\documentclass[10pt,a4paper]{article}
\usepackage[top=0.85in,footskip=0.75in]{geometry}

\usepackage{amsmath,amssymb}

\usepackage{changepage}

\usepackage[utf8x]{inputenc}

\usepackage{textcomp,marvosym}

\usepackage{cite}

\usepackage{nameref,hyperref}

\usepackage{microtype}
\DisableLigatures[f]{encoding = *, family = * }

\usepackage[table]{xcolor}

\usepackage{array}

\newcolumntype{+}{!{\vrule width 2pt}}

\newlength\savedwidth

\usepackage[hang,aboveskip=5pt,labelfont=bf,labelsep=period,justification=justified,singlelinecheck=off,textfont=small]{caption}
\renewcommand{\figurename}{Fig.}

\makeatletter
\renewcommand{\@biblabel}[1]{\quad#1.}
\makeatother

\date{}

\usepackage{lastpage,fancyhdr,graphicx}
\usepackage{epstopdf}
\fancyheadoffset[LR]{1.00in}
\fancyfootoffset[LR]{1.00in}

\newcommand{\kbT}{k_B T}

\renewcommand{\a}{{\bf a}}

\renewcommand{\c}{{\bf c}}
\newcommand{\e}{{\rm e}}

\newcommand{\ep}{\epsilon}

\renewcommand{\L}{{\cal L}}

\renewcommand{\a}{{\bf a}}

\newcommand{\flux}{\Delta_f}

\renewcommand{\i}{k}

\newcommand{\mutEffSecMom}{\mutEffSecMomSqrt^2}
\newcommand{\mutEffSecMomSqrt}{\epsilon_G}
\newcommand{\EQ}{\begin{equation}}
\newcommand{\EE}{\end{equation}}
\newcommand{\EQA}{\begin{eqnarray}}
\newcommand{\EEA}{\end{eqnarray}}

\newcommand{\cE}{{\cal E}}

\newcommand{\DG}{G} %
\newcommand{\sigtd}{\tilde{\sigma}}
\newcommand{\shape}{\mathcal C}
\newcommand{\C}{\shape}

\newcommand{\traitSecMom}{\epsilon_G}

\newcounter{firstbib}

\begin{document}
\vspace*{0.2in}

\begin{flushleft}
{\Large
\textbf\newline{Survival of the simplest: \\
the cost of complexity in microbial evolution} %
}
\newline
\\
Torsten Held\textsuperscript{1,\textbf{\P}},
Daniel Klemmer\textsuperscript{1,\textbf{\P}},
Michael L\"assig\textsuperscript{1,*},
\\
\bigskip
\textbf{1} Institut f\"ur Theoretische Physik, Universit\"at zu K\"oln,
Z\"ulpicherstr. 77, 50937, K\"oln, Germany
\\
\bigskip

\textbf{\P} These authors contributed equally to this work.

* mlaessig@uni-koeln.de

\end{flushleft}
\section*{Abstract}
The evolution of microbial and viral organisms often generates clonal interference, a mode of competition between genetic clades within a population. In this paper, we show that interference strongly constrains the genetic and phenotypic complexity of evolving systems. Our analysis uses biophysically grounded evolutionary models for an organism's quantitative molecular phenotypes, such as fold stability and enzymatic activity of genes. We find a generic mode of asexual evolution called {\em phenotypic interference} with strong implications for systems biology: it couples the stability and function of individual genes to the population's global speed of evolution. This mode occurs over a wide range of evolutionary parameters appropriate for microbial populations. It generates selection against genome complexity, because the fitness cost of mutations increases faster than linearly with the number of genes. Recombination can generate a distinct mode of sexual evolution that eliminates the superlinear cost. We show that positive selection can drive a transition from asexual to facultative sexual evolution, providing a specific, biophysically grounded scenario for the evolution of sex. In a broader context, our analysis suggests that the systems biology of microbial organisms is strongly intertwined with their mode of evolution. 

\section*{Introduction}

{A}sexually reproducing populations evolve under complete genetic
linkage. Hence, selection on an allele at one genomic locus can interfere with
the evolution of simultaneously present alleles throughout the genome. 
Linkage-induced interference interactions between loci include background
selection (the spread of a beneficial allele is impeded by linked deleterious
alleles), hitchhiking or genetic draft (a neutral or deleterious allele is
driven to fixation by a linked beneficial allele), and clonal interference
between beneficial alleles originating in disjoint genetic clades (only one of
which can reach fixation).
These interactions and their consequences for genome
evolution have been studied extensively in laboratory experiments
\cite{Wiser2013,Barroso-Batista2014}, natural populations~\cite{Betancourt2009,Strelkowa2012},
and theory~\cite{Gerrish1998,Desai01072007,Rouzine200824,Hallatschek01022011,schiffels2011emergent,Good2012,Neher08012013,Neher2013}.
The most prominent global effect of interference is to reduce the speed
of evolution, which has been observed in laboratory evolution experiments \cite{Visser1999,Perfeito2007,McDonald2016}.
The fitness cost of interference, which has also been measured \cite{Cooper2007,Couce2017},
is the center piece of classic arguments for the evolutionary advantage of sex
\cite{Fisher1930,Muller1932,Eigen1971,Felsenstein1974,Kondrashov1993}.
Much less clear is how interference affects the system-wide evolution of
molecular phenotypes, such as protein stabilities and affinities governing gene
regulation and cellular metabolism.
This is the topic of the present paper, which looks at the systems-biological consequences of interference evolution. We establish that interference generates a long-term degradation of an organism's molecular functions by the accumulation of deleterious mutations. This effect is strongly dependent on genome size: it becomes an evolutionary force constraining organismic complexity and driving the evolution of recombination.

\begin{figure*}[th!]
  \includegraphics[width=\textwidth]{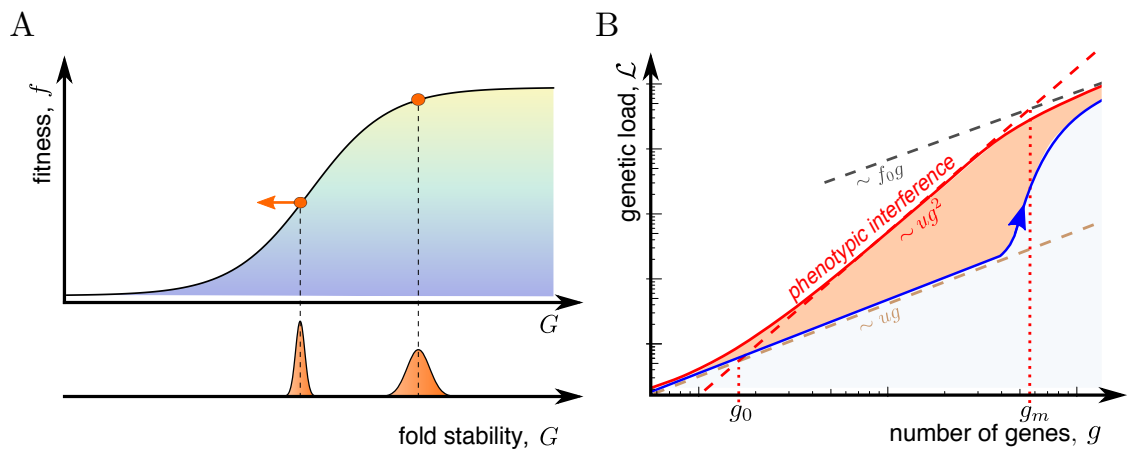}
\caption{ {\bf Fitness landscape and fitness cost of phenotypic interference.}
A:~Minimal biophysical fitness model. The fitness of an individual gene, $f$, is
a sigmoid function of its fold stability, $G$.
This function has a high-fitness region of stable, functional proteins (yellow),
an inflection point at intermediate fitness marking marginally functional
proteins, and a low-fitness region of dysfunctional proteins (blue). The
mutation-selection dynamics on this landscape generates high-fitness equilibria
($\sigtd \ll f_0$, red dot) and unstable states at lower fitness ($\sigtd \gtrsim
f_0$, red dot with arrow), depending on the fitness difference $f_0$ between functional and
dysfunctional proteins and the coalescence rate $\sigtd$; see also
Fig.~\ref{Fig:bio}B.
The fold stability distribution of a population in these states is shown
below.
B:~The total genetic load $\L$ in a genome is shown as a function of the number of genes, $g$, for different models of genome evolution.  Red line: Asexual evolution in the minimal biophysical model has an evolutionary regime of {\em phenotypic interference} where $\L$ increases quadratically with $g$; see Eq.~(\ref{sigtd})  and simulation data shown in Fig.~\ref{Fig:bio}A. This regime arises from the competition of phenotypic variants within a population. The nonlinear scaling of $\L$ sets in at a small gene number $g_0$ 
and ends at a much larger value $g_m$, which marks the crossover to genomes with a large fraction of dysfunctional genes (grey line). Blue line: Under asexual evolution in a model with discrete gene fitness effects, the onset of load nonlinearity and interference occurs at $g \sim g_m$ and is associated with the onset of Muller's ratchet \cite{Muller1964,Gordo2000,Rouzine200824}. {Brown} line: Sexual evolution reduces $\L$ to a linear function of $g$, if the recombination rate is above the transition point $R^*$ given by Eq.~(\ref{Rstar}). 
}
  \label{fig:1}
\end{figure*}

Our analysis is based on simple biophysical models of molecular
evolution~\cite{Gerland2002,Berg2004,Chen01102009,Goldstein2011,Serohijos201484,Manhart2015,
  Chi2016}.
In a minimal model, an individual's organism consists of $g$ genes and each gene carries a single  quantitative trait $G$, the stability of its protein.  
The trait is encoded in multiple sites of the gene sequence and is affected by mutations at these sites, most of which will make the protein less stable. Selection on a gene is described by a 
standard thermodynamic fitness landscape $f (G)$, which is a sigmoid function
with a high-fitness plateau corresponding to stable proteins and a low-fitness
plateau corresponding to  unfolded proteins (Fig.~\ref{fig:1}A). We also discuss
an stability-affinity protein model with a two-dimensional fitness landscape
$f(G,E)$; this model includes enzymatic or regulatory functions of genes,
specifically the protein binding affinity $E$ to a molecular target. The
genome-wide mutation-selection balance in these fitness landscapes
describes populations maintaining the functionality of their molecular traits;
we refer to this state as {\em housekeeping evolution}. We analyze its long-term evolutionary forces on genome architecture that arise independently of short-term adaptive processes, such as the evolution of resistance.

Over a wide range of model parameters, we find that housekeeping evolution takes
place in an evolutionary mode of {\em phenotypic interference}. In this mode, genetic
and phenotypic variants in multiple genes generate standing fitness variation
under complete genetic linkage, a so-called traveling fitness wave \cite{Desai01072007,Rouzine200824,Hallatschek01022011,Good2012,Neher08012013,Neher2013}. We show that phenotypic
interference is a system-wide collective dynamics with a universal feedback between the global fitness wave and selection on individual phenotypic variants. 
This feedback generates a fitness cost, defined as the difference between the mean population
fitness and the fitness maximum of fully functional genes, that increases
quadratically with $g$ (Fig. \ref{fig:1}B).
The fitness cost of interference quantifies its systems-biological effects: 
the maintenance of each gene degrades stability and function of all other genes by increasing the accumulation of deleterious mutations.
The cost nonlinearity sets in already at a small number of genes, $g_0$, and generates strong
selection against genome complexity in viable, asexually reproducing organisms.
This distinguishes our biophysical models from classical models of mutational
load, which predict a linear fitness cost up to a much larger error threshold $g_m$ associated with mutational meltdown \cite{Muller1964,Eigen1971,Gordo2000,Rouzine200824} (Fig.~\ref{fig:1}B). 

Remarkably, the genome-wide steady state of evolution affords an analytic
solution in our minimal model. We develop this solution in the following
section; then we turn to model extensions and biological consequences on genome
complexity under asexual evolution. Housekeeping evolution in these models also
provides a biophysically grounded rationale for the evolution of sex.
We show that long-term selective pressure on the recombination rate induces a first-order phase transition to a mode of sexual evolution without genome-wide interference, and we obtain a simple estimate of the transition recombination rate $R^{*}$ that can be directly compared to data.

\section*{Theory of phenotypic interference}

The solution of the minimal model has two parts that will be discussed in order.
First, the mean fitness variance of a single quantitative trait at evolutionary
equilibrium depends in a simple way on a global evolutionary parameter, the
coalescence rate $\sigtd$. Second, for the steady state of housekeeping evolution, the fitness variances of all traits combine to the total standing fitness variation, which in turn sets the coalescence rate and leads to a closure of the derivation. 

\subsubsection*{Evolution of a quantitative trait under interference selection}
The stability $G$ of a protein is the free energy difference between
the unfolded and the folded state (Methods). This trait gains heritable variation $\Delta_G$ by new mutations at a speed $u
\ep_G^2$, where $u$ is the total mutation rate and $\ep_G^2$ is the mean square
stability effect of its sequence sites. The trait loses variation by coalescence
at a rate $\sigtd $. These processes determine an equilibrium stability
variation  $\Delta_G = u \, \ep_G^2/ (2\sigtd)$. This type of relation is well
known for neutral sequence variation in models of genetic
draft~\cite{Gillespie909} and of traveling fitness
waves~\cite{Good2014,Rice01052015}.
It can
be derived more generally from a diffusion theory for quantitative traits under
selection; see \nameref{S1_Appendix} and
ref.~\cite{KlemmerLaessig2018}.
Next, we consider the mutation-selection equilibrium of a gene on the flank of the fitness landscape $f(G)$.  We equate the rate of stability increase by selection, 
$\Delta_G \, f' (G)$, with the rate of trait degradation by mutations,
$ u \ep_G$, using that most mutations in a functional trait are deleterious (Methods). 
This relates the mean square selection coefficient at trait sites, $s^2 = \ep_G^2 f'^2(G)$, and the fitness variance $\Delta_f\approx\Delta_G f'^2(G)$ to the coalescence rate, 
\EQ
s^2  =  4 \sigtd^2, 
\qquad
\Delta_f  =  2 u \sigtd. 
\label{deltaf}
\EE
These relations are universal; that is, they do not depend on details of the fitness
landscape and the trait effect distribution of sequence sites. Remarkably, trait fluctuations by genetic drift and genetic draft also leave their form invariant 
(\nameref{S1_Appendix} and \ref{S1_Fig}). 

Eqs.~(\ref{deltaf}) express a
salient feature of selection on quantitative traits: the strength of
selection on genetic variants is not fixed {\em a priori}, but is an emergent
property of the global evolutionary process. A faster pace of evolution, i.e.,
an increase in coalescence rate $\sigtd$, reduces the efficacy of selection
 \cite{schiffels2011emergent,Good2012,Rice01052015}. In
a downward curved part of
the fitness landscape, this drives the population to an equilibrium point of
lower fitness and higher fitness gradients. The resulting equilibrium tunes typical
selection coefficients to marginal relevance, where mean fixation times $1/s$
are of the order of the coalescence time $1/\sigtd$.
This point marks the crossover between effective neutrality ($s \ll \sigtd$) and
strong selection ($s \gg \sigtd$);
consistently, most but not all trait sites carry their beneficial
allele~\cite{schiffels2011emergent}.

\subsubsection*{Housekeeping evolution of multiple traits}
The equilibrium scenario of housekeeping evolution builds on the assumption that over long time scales, selection acts primarily to repair the deleterious effects of mutations, because these processes are continuous and affect the entire genome. In contrast, short-term adaptive processes are often environment-dependent, transient, and affect only specific genes. Here we discuss a closed solution of the phenotypic interference dynamics for housekeeping evolution. In \nameref{S1_Appendix}, we extend this approach to scenarios of adaptive evolution and show that these do not affect the conclusions of the paper. 

\begin{figure*}[th!]
  \includegraphics[width=\textwidth]{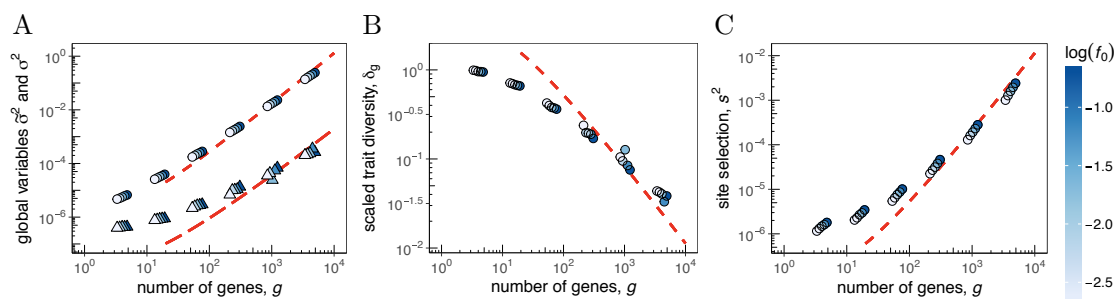}
  \caption{ 
    {\bf Global and local scaling under phenotypic interference.}
    A:~Average total fitness variance, $\sigma^2$ (circles) and coalescence rate $\sigtd^2$ (triangles) versus number of genes, $g$, for asexual evolution. Simulation data for different average gene selection coefficients $f_0$ (indicated by color) are compared to model results, $\sigma^2 \sim g^2/c$ (short-dashed line) and $\sigtd^2 \sim g^2/c^2$ (long-dashed line) %
     for $g > g_0 \sim 10^{2}$; Eqs.~(\ref{sigtd}) and~(\ref{c}). 
   B:~Average scaled trait diversity, $\delta_G = \Delta_G/\ep_G^2$, versus $g$. Simulation data (circles); model results, $\delta_G \sim c/g$ (dashed line; Eq.~(\ref{DeltaG})). Values $\delta_G < 1$ indicate that individual proteins are in the low-mutation regime. 
    C:~The mean square selection coefficient at sequence sites, $s^2$, versus $g$. Simulation results (circles);  model results, $s^2 = 
    4\sigtd^2 \sim g^2/c^2$ (short-dashed line as in A; Eq.~(\ref{c})). The scaling $s^2 \sim \sigtd^2$ is independent of $f_0$, signalling that site selection coefficients emerge from a feedback between global and local selection (see text). Other simulation parameters: $N=1000, u = 1.25\times10^{-3}$, $\ep_G / \kbT = 1$; see \nameref{S1_Appendix} for simulation details. 
  }
  \label{Fig:sim}
\end{figure*}

In a housekeeping equilibrium, the total fitness variation $\sigma^2$ is simply
the sum of the fitness variances of individual genes, $\sigma^2 = g \Delta_f$ (\ref{S2_Fig}).
Moreover, traveling wave theory shows that $\sigma^2$ and the 
coalescence rate $\sigtd$ are simply related, $\sigma^2 / \sigtd^2 = c_0 \log
(N \sigma) $, where $N$ is the population size and $c_0 \sim10^{2}$ \cite{Neher2013,Neher08012013}.  Together with Eq.~(\ref{deltaf}), we obtain the global fitness wave 
\EQ
{\sigma^2} = 4\frac{u^2 g^2}{c}, 
\qquad 
\sigtd  =2\frac{ug}{c},
\qquad 
\label{sigtd}
\EE
as well as corresponding characteristics of individual traits,
\EQ
\frac{\Delta_G}{\ep_G^2} = \frac{c}{4g}, 
\qquad 
{s^2} = 4\sigtd^2 = 16 \frac{u^2 g^2}{c^2}, 
\label{DeltaG}
\EE
in terms of the slowly varying parameter 
\EQ
c  = \frac{\sigma^2}{\sigtd^2} \approx {c_0} \log (N u g) .
\label{c}
\EE
As shown in Methods, this parameter has a simple interpretation: it estimates
the complexity of the fitness wave, that is, the average number of genes with simultaneously segregating
beneficial genetic variants destined for fixation. 
Fisher-Wright simulations of the minimal model confirm Eqs.~(\ref{sigtd}--\ref{c}); they reproduce the joint pattern of $\sigma^2$, $\sigtd^2$, $\Delta_G$, and $s^2$ and infer the wave complexity $c$~(Fig.~\ref{Fig:sim}).

These relations are the centerpiece of phenotypic interference theory. They show that the collective evolution of molecular quantitative traits under genetic linkage depends strongly on the number of genes that encode these traits. The dependence is generated by a feedback between the global fitness variation, $\sigma^2$, and mean square local selection coefficients at genomic sites, $s^2$. In \nameref{S1_Appendix}, we show that this feedback also tunes the evolutionary process to the crossover point between independently evolving genomic sites and strongly correlated fitness waves composed of multiple small-effect mutations.

\subsection*{Biological implications of phenotypic interference}

\subsubsection*{Interference selection against complexity}
\begin{figure*}[th!]
  \includegraphics[width=\textwidth]{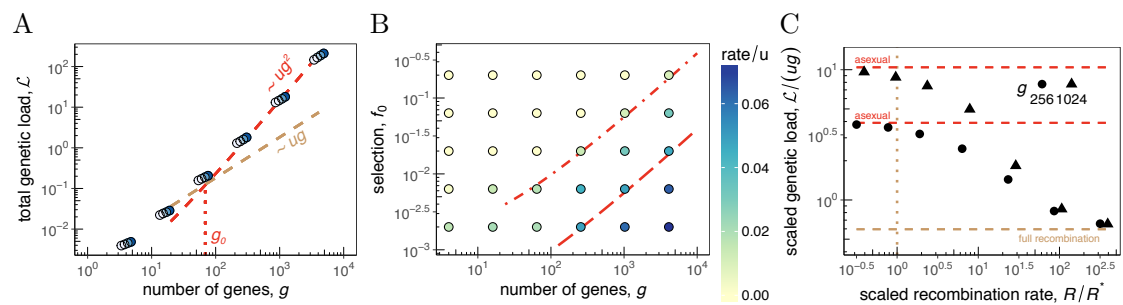}
  \caption{{\bf Genetic load, gene loss, and transition to sexual evolution.}
A:~Total genetic load $\L$  versus the number of genes $g$ for asexual evolution. Simulation results (circles) for different values of $f_0$ (indicated by color, as in Fig.~\ref{Fig:sim}); model results: interference load  $\L_{\rm int} \sim  ug^2/c$ ({red} line; Eq.~(\ref{load}))  for $g > g_0$ (dotted line) and null model $\L = ug$ ({brown} line) as in Fig.~\ref{fig:1}B. The superlinear behavior of $\L$ indicates strong selection against genome complexity. 
B:~Rate of gene loss (indicated by color, in units of $u$) as a function of the gene selection coefficient, $f_0$, and the number of co-evolving genes $g$. Genes with $f_0 \sim \sigtd $ (long-dashed line, cf. Fig.~\ref{Fig:sim}A) have appreciable loss rates; genes with $f_0 \gtrsim 10 \sigtd$ (dashed-dotted line) have negligible loss rates, i.e., are conserved under phenotypic interference. 
C:~Scaled genetic load, $\L/ (ug)$, versus scaled recombination rate, $R/R^*$, for different genome sizes. The observed load rapidly drops from the superlinear scaling of phenotypic interference, $\L = 4ug^2/c$ (asymptotic data: red lines), to the linear scaling of unlinked genes, $\L\sim ug$ ({brown} line). This signals a (fluctuation-rounded) transition to sexual evolution at the threshold recombination rate $R^* = 2ug/c$ (dotted line, see Eq.~(\ref{Rstar})). Other simulation parameters as in Fig.~\ref{Fig:sim};  see \nameref{S1_Appendix} for simulation details. 
 }
  \label{Fig:bio} 
\end{figure*}

The feedback of phenotypic interference has an immediate consequence for the
genetic load, which is determined by the average position of genes on the
fitness landscape. We first consider stable and functional genes located in the
concave part of the minimal model landscape $f(G)$ (Fig.~\ref{fig:1}A).
This part can be approximated by its exponential tail, where the load is
proportional to the slope $f'(G)$.
Eq.~(\ref{DeltaG}) then predicts a load $s\kbT/\ep_{G}  \approx
2  \sigtd$  per gene, where we have used that typical reduced effect sizes  $\ep_G/k_B T$ are of order~1 (Methods). This implies a superlinear scaling of the total equilibrium genetic load, 
\EQ
\L_{\rm int} (g) \approx 2  g \sigtd =
4\frac{ u g^2}{c}, 
\label{load}
\EE
which sets on at a small gene number $g_0$ given by the condition $g_0 \approx c/4$ (Fig.~\ref{fig:1}B, numerical
simulations are shown in Fig.~\ref{Fig:bio}A). The superlinearity of the genetic load is the most important biological effect of phenotypic interference. 
As detailed in \nameref{S1_Appendix} and \ref{S3_Fig}, this scaling holds more generally for a sufficient number of
quantitative traits evolving under genetic linkage; it does not depend on details of the
fitness landscape and of the underlying biophysical processes. For example, active protein degradation, a ubiquitous process that drives the thermodynamics of folding out of equilibrium~\cite{Hochstrasser1996}, does not affect our conclusions. Another example is the
stability-affinity model, which has two quantitative traits per gene that evolve
in a two-dimensional sigmoid fitness landscape $f(G,E)$~\cite{Manhart2015,Cheron2016}.
We show that under reasonable biophysical assumptions,
evolution in a stability-infinity model produces a 2-fold higher interference load than the minimal model,
$\L_{\rm int} (g) \approx 8 u g^2/c$.
Alternative models with a quadratic single-peak fitness landscape generate an even stronger nonlinearity of the load, $\L_{\rm int} (g) \sim g^3$. In contrast, a discrete model with a fitness effect $f_0$ of each gene shows a linear load up to a characteristic gene number $g_m = (f_0/u) \log (Nf_0)$ associated with the onset of mutational meltdown by Muller's ratchet~\cite{Muller1964,Gordo2000,Rouzine200824}. 

The interference load builds up with a time lag given by the relaxation time to equilibrium, 
\EQ
\tau = \frac{1}{u} = 2\frac{g}{c} \, \frac{1}{\sigtd}. 
\label{tau}
\EE
Deleterious mutations in an organism's genes build up on a time scale $\tau$, which exceeds the coalescence time $\sigtd^{-1}$.
Therefore, the load $\L_{\rm int}$ affects the long-term fitness of a population
against competing lineages. Specifically, it generates strong long-term
selection against genome complexity: the fitness cost for each additional gene,
$\L'_{\rm int} (g)$, can take sizeable values even at moderate genome size.
For example, in a ``standard'' microbe of the complexity of {\em E. coli}, a
10\,\% increase in gene number may incur an additional load $\Delta \L \approx
3 \times 10^{-2}$  under the stability-affinity model (with parameters $g = 5000, u =
10^{-6}, N = 10^{8}$). In comparison, the discrete model leads to a much smaller
value $\Delta \mathcal L = 5 \times 10^{-4}$ for the same parameters.

It is instructive to compare the interference load of an extra gene with its
physiological fitness cost  $\L'_{\rm phys} (g)$, which is generated primarily
by the synthesis of additional proteins (and is part of the overall fitness
effect $f_0$). For a gene with an average expression level, $\L'_{\rm phys} (g)
= \lambda/g$ with a constant $\lambda \sim 1$ reflecting the (re-)allocation of metabolic
resources in the cell; see refs.~\cite{Scott2010,Lynch2015}.
This cost acts as a selective force on changes of genome size,
which take place within a coalescence interval $\sigtd^{-1}$.
Importantly, $\L'_{\rm phys}$ is much smaller than $\L'_{\rm int}$ for a standard microbe,
suggesting a two-scale evolution of genome sizes.
On short time scales, the dynamics of gene numbers is permissive and allows the
rapid acquisition of adaptive genes; these changes are neutral with respect to
$\L'_{\rm int}$.
On longer time scales (of order $\tau$), marginally relevant genes are pruned in a more stringent way,
for example, by invasion of strains with more compact genomes. 

\subsubsection*{Interference drives gene loss}
The near-neutral dynamics of genome size extends to gene losses, which become likely when a gene gets close to the inflection point of the sigmoid fitness landscape (Fig.~\ref{fig:1}A). The relevant threshold gene fitness, $f_0^c$, is set by the coalescence rate, which leads to 
\EQ
f_0^c \sim 2 \sigtd = \frac{4 ug}{c}
\label{eq:f0crit}
\EE
in the minimal model. Strongly selected genes ($f_0 \gg 2 \sigtd$) have
equilibrium trait values firmly on the concave part of the landscape, resulting
in small loss rates of order $u \exp(- f_0 / 2 \sigtd)$; these genes can be
maintained over extended evolutionary periods. Marginally selected genes ($f_0
\lesssim 2 \sigtd$) have near-neutral loss rates of order $u$
\cite{schiffels2011emergent}, generating a continuous turnover of genes.
According to Eq.~(\ref{eq:f0crit}), the threshold $f_0^c$ for gene loss
increases with genome size, which expresses again the evolutionary constraint on
genome complexity.
The dependence of the gene loss rate on $f_0$ and $\sigtd$ is confirmed by simulations (Fig.~\ref{Fig:bio}B). The housekeeping coalescence rate $\sigtd = 2 ug/c$  sets a lower bound for the fitness threshold $f_0^c$, adaptive evolution can lead to much larger values of $\sigtd$ and $f_0^c$.

\subsubsection*{The transition to sexual evolution}
Recombination breaks up genetic linkage at a rate $R$ per genome and per generation ($R$ is also called the genetic map length). Evolutionary models show that recombination generates linkage blocks that are units of selection; a block contains an average number $\xi$ of genes, such that there is one recombination event per block and per coalescence time, $R \xi /(g \sigtd (\xi)) = 1$~\cite{Weissman2012,Neher2013a,Neher2013}. Depending on the recombination rate, these models predict a regime of asexual evolution, where selection acts on entire genotypes  ($\xi \sim g$), and a distinct regime of sexual evolution with selection acting on individual alleles ($\xi \ll g$).  Here we focus on the evolution of the recombination rate itself and establish a selective avenue for the transition between asexual and sexual evolution. With the phenotypic interference scaling $\sigtd (\xi) = 2u \xi /c$ for $\xi \gtrsim c$, as given by Eq.~(\ref{sigtd}), our minimal model produces an instability at a threshold recombination rate
\EQ
R^* = \frac{2ug}{c}. 
\label{Rstar}
\EE
This signals a first-order phase transition to sexual evolution with the genetic load as order parameter (Fig.~\ref{Fig:bio}C). For $R < R^*$, the population is in the asexual mode of evolution ($\xi \sim g$), where interference produces a superlinear load $\L_{\rm int} = 2ug^2/c$. For $R > R^*$, efficient sexual evolution generates much smaller block sizes ($\xi \sim c$). In this regime, the mutational load drops to the linear form $\L = ug \ll \L_{\rm int}$, providing a net long-term fitness gain $\Delta \L \simeq \L_{\rm int}$. However, the process of recombination itself entails a direct short-term cost $\L_{\rm rec}$~\cite{Kondrashov1993}. If we assume that cost to be of order~1 per event, we obtain $\L_{\rm rec} \sim R^* = \sigtd$ close to the transition. This cost is much smaller than the gain $\Delta \L$ and remains marginal (i.e., $\L_{\rm rec}/ \sigtd \sim 1$). 

Together, our theory of phenotypic interference suggests a specific two-step scenario for the evolution of sex. Recombination at a rate of order $R^*$ is near-neutral at short time scales, so a recombining variant of rate $R^*$ arising in an asexual background population can fix by genetic drift and draft. Recombining strains acquire a long-term benefit $\Delta \L \sim gR^* = g \sigtd$, so they can outcompete asexual strains in the same ecological niche. The threshold rate $R^*$ is of the order of the genome-wide mutation rate $ug$, so even rare facultative recombination can induce the transition. This scenario builds on the basic biophysics of molecular traits but does not require {\em ad hoc} assumptions on adaptive pressure, on rate and effects of beneficial and deleterious mutations, or on genome-wide epistasis~\cite{Kondrashov1993}. It is at least consistent with observed recombination rates in different parts
of the tree of life: genome average values are always well above $R^*$; a
high-resolution recombination map of the {\em Drosophila} genome shows
low-recombining regions with values above but of order $R^*$~\cite{Comeron2012,Schiffels2017} (\ref{S1_Table}).

\section*{Discussion}
Here we have developed the evolutionary genetics of multiple quantitative traits
in non-recombining populations.
We find a specific evolutionary mode of phenotypic interference,
which is characterized by a feedback between global fitness variation and local
selection coefficients at genomic sites.
This feedback generates highly universal features, which include
the complexity of the evolutionary process and the scaling of
coalescence rate and genetic load with gene number,
as given by Eqs. (\ref{sigtd})--(\ref{c}). 

Phenotypic interference produces strong selection against genome complexity in
asexual populations, which implies selection in favor of 
recombination above a threshold rate $R^*$ given by Eq.~(\ref{Rstar}).
The underlying genetic load originates from the micro-evolutionary interference
of phenotypic variants within a population and unfolds with a time delay beyond
the coalescence time, as given by Eq.~(\ref{tau}).
Therefore, the interference load is a macro-evolutionary selective force that
impacts the long-term fitness and survival of a population in its ecological
niche.

Molecular complexity, the broad target of phenotypic interference, can be regarded as a key systems-biological observable. In our simple biophysical models, we measure complexity by number of stability and binding affinity traits in a proteome. More generally, we can define complexity as the number of (approximately) independent molecular quantitative traits, which includes contributions from an organism's regulatory, signalling, and metabolic networks that scale in a nonlinear way with genome size. Interference selection affects the complexity and architecture of all of these networks, establishing new links between evolutionary and systems biology to be explored in future work. 

\section*{Methods}

{ 
\subsection*{Biophysical fitness models}
In thermodynamic equilibrium at temperature $T$, a protein is folded with probability $p_+ (G) = 1/[1 + \exp(- \DG/k_BT)]$, where $\DG$ is the Gibbs free energy difference between the unfolded and the folded state and $k_B$ is Boltzmann's constant. A minimal biophysical fitness model for proteins takes the form 
\EQ
f(G) = f_0 \, p_+ (G) = \frac{f_0}{1 + \exp(- \DG/k_BT)}
\label{eq:fG}
\EE
with a single selection coefficient capturing functional benefits of folded proteins and metabolic costs of misfolding~\cite{Chen01102009,Goldstein2011,Serohijos201484}. Similar fitness models based on binding affinity have been derived for transcriptional regulation~\cite{Gerland2002,Berg2004,Mustonen2008}; the rationale of biophysical fitness models has been reviewed in refs.~\cite{Lassig2007,Chi2016}. In \nameref{S1_Appendix}, we introduce alternative fitness landscapes for proteins and show that our results depend only on broad characteristics of these landscapes. The minimal global fitness landscape for a system of $g$ genes with traits $G_1, \dots, G_g$ and selection coefficients $f_{0,1}, \dots, f_{0,g}$ is taken to be additive, i.e., without epistasis between genes, 
\EQ
f(\DG_1,\dots,\DG_g) = \sum_{i=1}^g \frac{f_{0,i} }{ [1 + \exp(- \DG_i/k_BT)]}. 
\EE

\subsection*{Evolutionary model}
We characterize the population genetics of an individual trait $G$ by its population mean $\Gamma$ and its expected variance $\Delta_G$. The trait mean follows the stochastic evolution equation 
\EQ
\dot \Gamma = - u \kappa \traitSecMom + \Delta_{G} f'(\Gamma) + \chi( t)
\label{Gammadot}
\EE
with white noise $\chi(t)$ of mean $\langle \chi(t) \rangle = 0$ and variance $\langle \chi(t) \chi(t') \rangle = \sigtd \Delta_{G} \, \delta(t - t')$. This dynamics is determined by the rate $u$, the mean effect $(- \kappa) \ep_G$, and the mean square effect $\ep_G^2$ of trait-changing mutations, which determine the diversity $\Delta_{G} = u \ep_G^2 / (2\sigtd)$~\cite{Nourmohammad:2013ty, KlemmerLaessig2018}. We use effects $\traitSecMom\approx 1\text{--}3 \,\kbT$, which have been measured for fold stability \cite{Tokuriki2007,Zeldovich07} and for molecular binding traits ~\cite{Gerland2002,Kinney2010,Tugrul2015}, and a mutational bias $\kappa =1$, which is consistent with the observation that most mutations affecting a functional trait are deleterious. 

\subsection*{Housekeeping equilibrium} The deterministic equilibrium solution ($\dot \Gamma = 0, \chi = 0$) of Eq.~(\ref{Gammadot}) determines the dependence of $\Delta_G$ and the associated fitness variance $\Delta_f = \Delta_G f'^2 (G)$ on~$\sigtd$, as given by Eq.~(\ref{deltaf}); the same scaling follows from the full stochastic equation (\nameref{S1_Appendix}). The derivation of the global housekeeping equilibrium,  Eqs.~(\ref{sigtd})--(\ref{c}), uses two additional inputs: the additivity of the fitness variance, $\sigma^2 = g \Delta_f$, which is confirmed by our simulations (\ref{S2_Fig}), and the universal relation $\sigma^2 / \sigtd^2 = c_0 \log (N \sigma) $~\cite{Neher2013,Neher08012013} in a travelling fitness wave, where the coalescence rate $\sigtd$ is generated predominantly by genetic draft. Eqs.~(\ref{sigtd})--(\ref{c}) determine further important characteristics of phenotypic interference: 
\begin{enumerate}
\item[(a)] The complexity of the fitness wave, defined as the average number of beneficial substitutions per coalescence time, is $(v_+ g)/ \sigtd \sim ug / 2 \sigtd = c/4$, using that trait-changing mutations are marginally selected, Eq.~(\ref{deltaf}), and have nearly neutral fixation rates $v_{+} \sim u/2$ per gene. 
\item[(b)] The evolutionary equilibria of stable genes ($f_0 \gg \sigtd$) are located
in  the high-fitness part of the minimal fitness landscape, $f \simeq f_0 [1 -
\exp(- \DG/k_BT)]$.
These genes have an average fitness slope $f' = (\Delta_f / \Delta_G)^{1/2} = 2\sigtd/\ep_G$, an average trait $\Gamma = - \kbT \log (2 \sigtd \kbT / f_0 \traitSecMom)$, and an average load $\L_{\rm int} (g)$ given by Eq.~(\ref{load}). 
\item[(c)] The scaling regime of Eqs.~(\ref{sigtd})--(\ref{c}) sets in at a gene number $g_0$ given by the condition $g_0 = c/4$; this point also marks the crossover from the linear load $\L_0 (g) = ug$ to the nonlinear form $\L_{\rm int} (g)$. 
\end{enumerate}

\section*{Acknowledgments}
We thank T.~Bollenbach for discussions. This work has been supported by Deutsche Forschungsgemeinschaft grants SFB 680 and SFB 1310 (to ML).  We acknowledge computational support by the CHEOPS platform at University of Cologne.

\newpage
\renewcommand{\figurename}{} 
\renewcommand{\thefigure}{S\arabic{figure} Fig}
\setcounter{figure}{0}
\renewcommand{\tablename}{} 
\renewcommand{\thetable}{S\arabic{table} Tab}
\setcounter{table}{0}

\part*{S1 Appendix}
\label{S1_Appendix}

\section{Trait diversity and cross-over scaling of the fitness wave}

\paragraph{Equilibrium of trait diversity.}

Consider a quantitative trait $G$ evolving by mutations, coalescence caused by
genetic drift and genetic draft, and stabilizing selection in a fitness
landscape $f(G)$.
Mutations and coalescence alone generate an equilibrium of the trait diversity $\Delta$, 
\EQ
  { \Delta_G}  = \frac{u{\epsilon_G^2} }{2\sigtd}, 
  \label{eq:DeltaInterference}
\EE
as derived in the main text and refs.~\cite{Nourmohammad:2013ty, KlemmerLaessig2018}. This expression is valid if stabilizing selection on the trait diversity can be neglected, i.e., if~\cite{Nourmohammad:2013ty}
\EQ
\frac{\L_\Delta}{\sigtd} \equiv \frac{\Delta_{G}|f''(\Gamma)|}{\sigtd} \lesssim 1. 
\label{cond}
\EE
Here we show that this condition is self-consistently fulfilled throughout the
phenotypic interference regime. Evaluating the expected fitness curvature in the
high-fitness part of the minimal fitness landscape, Eq.~(\ref{eq:fG}),
where $f'' (\Gamma) = - f'(\Gamma)/k_BT$, and in the mutation-coalescence equilibrium given by Eq.~(\ref{eq:DeltaInterference}), we obtain $f'' = -2\sigtd/ (\epsilon_g k_B T)$. By Eq.~(\ref{DeltaG}), the condition~(\ref{cond}) then reduces to 
\EQ
\frac{\Delta_G}{\epsilon_G^2} = \frac{c}{4g} \lesssim 1,
\EE
which is identical to the condition for phenotypic interference given in the main text. This relation expresses an important scaling property of the phenotypic interference regime: individual traits evolve in the low-mutation regime and are monomorphic at most times. In contrast,  the the global trait diversity defines a polymorphic fitness wave, 
\EQ
\frac{4g  { \Delta_G}}{{\epsilon_G^2}} = \frac{\sigma^2 }{ \sigtd^2 }= c \gtrsim g_0. 
\EE

\paragraph{Cross-over scaling of the fitness wave.}
A travelling fitness wave maintained by mutations at genomic sites with a fixed selection coefficient $s$ has two distinct scaling regimes~\cite{Neher2013,Neher2013a}, 
\begin{equation}
  \sigma^2 = 
    \begin{cases}
      sug, & (g \lesssim g_c)
      \\
      c \sigtd^2 = (\frac{c}{4})^{1/3}(s^{2} ug)^{2/3} 
      & (g \gtrsim g_c),
    \end{cases}
\end{equation}
which correspond to independently evolving sites and to an asymptotic fitness
wave with strong interference selection, respectively.
At the crossover point $g_c = c s/ (4u)$, the relation 
\EQ
\sigtd (g_c) = \frac{2u g_c}{c} = \frac{1}{2}s 
\EE
is valid. Comparing this relation with the {\em generic} scaling under phenotypic
interference, $\sigtd = 2 ug / c = s/2$ as given by Eqs.~(\ref{sigtd})
and~(\ref{DeltaG}), we conclude that the phenotypic fitness wave is locked in
the crossover region of marginal interference.
As discussed in the main text, this feature reflects the feedback between global
and local selection in a phenotypic fitness landscape, which tunes selection
coefficents to the $g$-dependent value $s = 4 ug / c$.
Consistently, the phenotypic fitness wave has a fitness variance $\sigma^2 \sim g^2$, compared to the scaling $\sigma^2 \sim g^{4/3}$ of the asymptotic regime at fixed selection coefficients (up to log corrections).

\section{Stochastic theory of phenotypic interference}
In the main text, we derive the scaling relations of phenotypic interference, Eqs.~(\ref{deltaf}) -- (\ref{load}), using the evolution equation for quantitative traits, Eq.~(\ref{Gammadot}), in its deterministic limit ($\chi = 0$). Here we show that the full evolution equation generates the same scaling. We convert Eq.~(\ref{Gammadot}) into an equivalent diffusion equation~\cite{Nourmohammad:2013ty,KlemmerLaessig2018} for the probability density $Q(\Gamma, t)$, 
\EQ
\frac{\partial}{\partial t} \, Q (\Gamma, t) = 
\left [ \sigtd \Delta_{G} \frac{\partial^2}{\partial \Gamma^2} +
  \frac{\partial}{\partial \Gamma} \left ( \kappa \epsilon_G u - \Delta_{G} f'(\Gamma) \right )
\right ] \,  Q (\Gamma, t) 
\EE
with the average trait diversity $\Delta_G$ given by Eq.~(\ref{eq:DeltaInterference}). The equilibrium probability distribution $Q_{\rm eq} (\Gamma)$ describes the stationary fluctuations of the population mean trait $\Gamma (t)$ of a stable gene around its long-term average $\langle \Gamma \rangle = \int \Gamma \, Q_{\rm eq} (\Gamma) \, d\Gamma$; these fluctuations are generated by genetic drift and (predominantly) genetic draft. In the biophysical fitness landscape, Eq.~(\ref{eq:fG}), the equilibrium distribution can be evaluated analytically, 
\EQ
Q_{\rm eq} (\Gamma) = \frac{ \left(\frac{f_0}{\sigtd }\right)^{2\kappa\frac{\kbT}{\mutEffSecMomSqrt} } \exp\left(-  \frac{2\kappa\Gamma}{\mutEffSecMomSqrt} -\frac{f_0}{\sigtd } \e^{- \Gamma /\kbT}\right)}{\kbT{\,\text{Gamma} \left(2\kappa \frac{\kbT}{\mutEffSecMomSqrt}  \right)}} ,
\label{QGamma}  
\EE
where Gamma and PolyGamma are standard transcendental functions. This function is plotted in Fig.~A in \ref{S1_Fig}. The resulting average, 
\EQ
\frac{\langle \Gamma \rangle}{\kbT} = - \log \left(\frac{\sigtd }{f_{0}  } \right)  -{\rm PolyGamma}\left( 2\kappa\frac{\kbT}{\mutEffSecMomSqrt} \right),
\label{Gamma_ave}
\EE
shows that genes are slightly more stable than estimated from the deterministic average derived in the main text, $\Gamma  / \kbT =  - \log (2 \kappa \sigtd \kbT / f_0 \epsilon_G)$. Through the nonlinearity of the fitness landscape, the fluctuations of the mean trait $\Gamma$ induce fluctuations of the conditional average fitness variance, $\langle \Delta_f \rangle (\Gamma) = \Delta_G f'^2 (\Gamma)$. We obtain the equilibrium distribution
\EQ
 Q_{\rm eq} (\flux) = \text{Gamma}_{\rm gen}\left(\flux; \;2\kappa\frac{\kbT}{\mutEffSecMomSqrt}, \frac{1}{2}u\sigtd \frac{\mutEffSecMom}{(\kbT)^{2}}, \frac{1}{2}, 0\right) , 
\label{QDiv}
\EE
with  $\text{Gamma}_{\rm gen}$ denoting the generalized gamma distribution (Fig.~B in \ref{S1_Fig}). The average fitness variance 
\EQ
 \langle\flux\rangle
= 2\sigtd  u \kappa^2 \left(1+ \frac{\mutEffSecMomSqrt}{ 2\kappa\kbT}\right)  
\label{Delta_ave}
\EE
differs from its deterministic counterpart, Eq.~(\ref{deltaf}) by a prefactor of order 1. Similarly, the $\Gamma$ fluctuations induce fluctuations of the interference load of individual genes, 
\EQ
Q_{\rm eq} (\L_{\rm gene}) = \text{Gamma}_{\rm dist}\left(\L_{\rm gene}; \;2\kappa\frac{\kbT}{\mutEffSecMomSqrt},\sigtd\right)  
\label{Qf}
\EE 
(Fig.~C in \ref{S1_Fig}). The resulting dependence 
\EQ
  \langle \L_{\rm gene} \rangle = 2\kappa \frac{k_B T}{\mutEffSecMomSqrt} \sigtd  
\label{ave}
\EE
is identical to the deterministic case; the fluctuation effect on $\langle \Gamma\rangle$, Eq.~(\ref{Gamma_ave}), is offset by the fluctuation load in a downward-curved fitness landscape.

\section{Model extensions}
In this section, we develop alternative evolutionary models of quantitative traits under genetic linkage. The mode of phenotypic interference, which is characterized by a superlinear scaling of the genetic load with genome complexity, occurs in all cases, suggesting it is a generic property of this class of models. Specifically, we discuss housekeeping dynamics in extended models of protein evolution and we extend our analysis to adaptive processes. 

\paragraph{Active protein degradation.} 
This non-equilibrium process affects a wide range of proteins, for example
through the ubiquitin-proteasome pathway\cite{Hochstrasser1996}. It ensures that
regulatory proteins are rapidly cleared once their function ends (at a
particular point of the cell cycle).
Consider a simple model, which has a constant rate $K^-$ of active degradation and a rate $K_{G}^{+} = K_G^{0} \e^{G/k_B T}$ for the folding process. Here we do not model details of the pathways of protein synthesis from and degradation into amino acid constituents, which would only affect the total protein concentration but not their state probabilities. In the steady state, proteins are folded with probability 
\EQ
\tilde p_+ (G) = \frac{1}{1 + \nu_{G}\e^{- \DG/k_BT}},
\label{tildep}
\EE
where $\nu_{G} = K^{-}/K_G^{0}$.
Hence, this model retains the sigmoid form of the fitness landscape given in Eq.~(9) and shown in Fig.~1 of the main text, and our evolutionary conclusions remain invariant. 

\paragraph{Stability-affinity model. } 
This model extends the minimal protein model discussed in the main text by
explicitly including protein function, which is assumed to be mediated through
binding to a molecular target. Proteins can be in three thermodynamic states:
functional, i.e., folded and target-bound ($++$), folded and unbound ($+-$), and
unfolded ($- -$). We assume that unfolded proteins cannot bind their target, which implies that the fourth state of unfolded proteins localized to their target ($-+$) is suppressed by  the entropy loss of localization. 

We consider two different thermodynamic ensembles of these proteins. In {\em thermodynamic equilibrium,} the statistics of this ensemble is governed by two quantitative traits,
which are defined as free energy differences: the fold stability $G \equiv
G_{--} - G_{+-}$ and the reduced binding affinity $E \equiv G_{+-} - G_{++}$, which includes the entropy loss of  localization and depends on the ligand concentration. The equilibrium state probabilities $p_{++}$, $p_{--}$, and
$p_{--}$ are given by  Boltzmann statistics depending on the traits $G$ and
$E$; in particular, 
\EQ
p_{++} (G,E) = \dfrac{1}{1 +  \e^{-E/\kbT} + \e^{-(E+G)/\kbT}}.
\label{p++1}
\EE
Equilibrium models of this kind are well known in protein biophysics~\cite{Phillips2013,Monod1965}, and have been used to build fitness landscapes~\cite{Manhart2015,Cheron2016}. {\em Active degradation} is again a ubiquitous process that drives the thermodynamics out of equilibrium; this process is particularly relevant for target-bound proteins that would have a long lifetime at thermodynamic equilibrium. Here we assume a single degradation rate $K^-$ for the processes  $(++)\rightarrow (--)$ and $(+-)\rightarrow (--)$, a rate $K_G^{+} = K_G^{0} \e^{G/k_B T}$ for the folding process $(--)\rightarrow (+-)$, and a rate $K_E^{+} = K_E^0 \e^{E/k_B T}$ for the binding process $(+-)\rightarrow (++)$. In this model, the folding and binding processes decouple, and we obtain  %
the non-equilibrium steady-state probability 
\EQ
p_{++}(G,E) =   \frac{1}{\big (1 + \nu_{G} \e^{- \DG/k_BT} \big ) \big (1 + (1+\nu_{E}) \e^{-E/\kbT} \big )}
\EE
with $\nu_{G} = K^{-}/K_G^{0}$ and  $\nu_{E} = K^{-}/K_E^{0}$. From these ensembles, we build thermodynamic fitness landscapes 
\EQ
f(G,E) =  f_0 \, p_{++} (G,E)
\label{fEG}
\EE
analogous to Eq.~(\ref{eq:fG}); these landscapes are plotted in Fig.~A and~B in \ref{S3_Fig}. 

The population genetics of the two-trait system is described by the population mean values $\Gamma_G$ and $\Gamma_E$, the diversities $\Delta_{GG}$ and $\Delta_{EE}$, and the covariance $\Delta_{GE}$. Under mutations, coalescence, and selection given by the fitness landscape $f(G,E)$, the mean traits follow a stochastic evolution equation analogous to Eq.~(\ref{Gammadot}), 
\EQ
\begin{pmatrix} \dot \Gamma_G \\ \dot \Gamma_E \end{pmatrix} = 
-\begin{pmatrix} u_{G}\kappa_{G} \epsilon_{G} \\ u_{E}\kappa_{E}\epsilon_{E}\end{pmatrix}
+
\begin{pmatrix}
  \Delta_{GG} & \Delta_{GE} \\
  \Delta_{GE} & \Delta_{EE}
\end{pmatrix}
\begin{pmatrix}
  \partial_G f(E,G)\\
  \partial_E f(E,G)
\end{pmatrix}
+
\begin{pmatrix} \chi_G \\ \chi_E \end{pmatrix}
\EE
with white noise of mean and variance 
\EQ
\begin{pmatrix} \langle \chi_G \rangle  \\ \langle \chi_E
  \rangle \end{pmatrix}  =
\begin{pmatrix}
  0 \\
  0
\end{pmatrix}
,
\qquad
\begin{pmatrix} \langle \chi_G (t) \chi_G(t')  \rangle  & \langle \chi_G (t) \chi_E(t') \rangle
\\ 
\langle \chi_G (t) \chi_E(t') \rangle & \langle \chi_E (t) \chi_E(t') \rangle 
\end{pmatrix}  = \sigtd \delta(t - t')
\begin{pmatrix}
  \Delta_{GG} & \Delta_{GE} \\
  \Delta_{GE} & \Delta_{EE}
\end{pmatrix}.
\EE
Here we discuss the simplest stationary states of housekeeping evolution in this model, using the deterministic limit of the evolution equation ($\chi_G = \chi_E = 0$). The trait diversities $\Delta_{GG}$ and $\Delta_{EE}$ are given as in Eq.~(\ref{eq:DeltaInterference}), and we assume that pleiotropic sites have uncorrelated effects on both traits, i.e., $\Delta_{GE} = 0$; this has  recently been observed in \cite{Otwinowski2018}. As in the main text, we set $\kappa_G = \kappa_E = 1$, which says that most random mutations reduce stability and affinity. 

In equilibrium, the high-fitness part of the fitness landscape takes the asymptotic form 
 $f(G,E) \simeq f_{0} [1 -\e^{-E/\kbT}(1+\e^{-G/\kbT}+\e^{-E/\kbT} )] +
 O((e^{-E/\kbT} , e^{-G/\kbT})^3)$.
 The mutation-selection equilibrium leads to mean trait values
 \EQ
 \begin{pmatrix}
   \Gamma_G\\
   \Gamma_E
 \end{pmatrix}
 \approx
 \begin{pmatrix}
  k_B T\log\left(\frac{\epsilon_G}{\epsilon_E} -1\right)\\
   - k_B T\log \left (2 \frac{\sigtd}{f_0} \left(\frac{\kbT}{\epsilon_E} -
     \frac{\kbT}{\epsilon_G}\right)\right )
 \end{pmatrix},
 \EE
where only the $E$-component depends on the coalescence rate $\sigtd$. Comparison with the minimal model, Eq.~(1), shows that 
in the stable part of the fitness landscape, the equilibrium stability-affinity model becomes an essentially one-dimensional problem for the affinity trait $E$~\cite{Manhart2015}.  
The total fitness variance per gene, $\Delta_f = 2(u_G + u_E) \sigtd$, is of the universal form~[1] with an effective mutation rate
\EQ
u = u_G + u_E.
\label{utot}
\EE
We conclude that housekeeping evolution in this model follows the same scaling as in the minimal model, Eqs.~(\ref{deltaf}) -- (\ref{Rstar}), with the parameter $u$ given by Eq.~(\ref{utot}). However, the equilibrium model lacks evolutionary stability, because lack of folding stability ($G>0$) can be compensated by a stronger binding affinity.  

With active degradation, the high-fitness part of the fitness landscape takes the asymptotic form 
 $ f(G,E) \simeq f_{0}[1-(1+\nu_{E})\e^{-E/\kbT}-\nu_{G}\e^{-G/\kbT}] + O(
(e^{-E/\kbT} , e^{-G/\kbT})^2 )$.
Hence, for stable genes ($f_0 \gg \sigtd$), the evolutionary dynamics of the traits $G$ and $E$ becomes approximately independent. The traits of each gene are at a mutation-selection equilibrium of the universal form~(\ref{deltaf}), generating a combined fitness variance $\Delta_f = 2(u_G + u_E) \sigtd$. Therefore, housekeeping evolution in this model also follows the same scaling as in the minimal model, Eqs.~(\ref{deltaf}) -- (\ref{Rstar}), with a total mutation rate per gene given by Eq.~(\ref{utot}) and an effective value of $g$ that is twice the number of genes, 
\EQ
g_{\rm eff} = 2g. 
\EE
In particular, the system-wide interference load is about twice the value of the minimal model, $\L_{\rm int} \approx 8ug^2 / c$, as used in the main text. This estimate disregards the additional contribution from the enhanced total mutation rate, Eq.~(\ref{utot}), which takes into account that $u_E \ll u_G$ for many binding domains. 

The form invariance of housekeeping evolution in these models shows the robustness of the phenotypic interference mode. It also suggests that in more general contexts, we can define genomic complexity as the number of quantitative traits that evolve (approximately) independently; see the Discussion of the main text.

\paragraph{Single-peak fitness model.} 
A minimal model of stabilizing selection is a quadratic 
landscape \cite{Lande1976,deVladar2014,nourmohammad2013universality}, 
\begin{align}
f(E) =
-f_0(E-E^*)^2 
\label{eq:fquad}
\end{align}
(Fig.~C in \ref{S3_Fig}). This model penalizes deviations from an optimal trait value $E^{*}$. In contrast to the biophysical landscape, there is no gene loss in a quadratic landscape, because there are no constraints on its slope. As long as mutations generate trait equilibria predominantly on one flank of the landscape, the basic scaling of phenotypic interference, Eqs.~(\ref{deltaf}) -- (\ref{c}), is universal and, hence, the same as in the minimal model. The genetic load  for a single gene, $\L_{\rm gene} = -f(\Gamma) = u^2 \epsilon_E^2 / 4\Delta_E^2 f_0$,  has been derived in \cite{Nourmohammad:2013ty}. With $\Delta_E$ given by Eq.~(\ref{eq:DeltaInterference}), we find a system-wide interference load 
\begin{equation}
   \L_{\rm int} =  g \L_{\rm gene}  = \frac{4u^{2} g^3}{c^{2}\epsilon_E^2f_0}.
\end{equation}
 Hence, the single-peak model has an even stronger load nonlinearity than the biophysical fitness landscapes.

\paragraph{Phenotypic interference in adaptive evolution.}
Here we show that the phenotypic interference scaling extends to simple models of adaptive evolution. In the minimal biophysical model, we assume that protein stabilities are still at an evolutionary equilibrium of the universal form~(\ref{deltaf}), generating a combined fitness variance $g \Delta_f = 2 g u \sigtd$. However, the global fitness variance acquires an additional contribution from adaptive evolution of other system functions, 
\begin{align}
\sigma^{2} =  c \sigtd^2 = 2 \sigtd  ug + \phi, 
\end{align}
where $\phi$ is the {\em fitness flux} or rate of adaptive fitness gain~\cite{Mustonen2010}. Mathematically, this term quantifies the deviations of the adaptive evolutionary  process from equilibrium (defined by detailed balance). Closure of the modified dynamics leads to an increased coalescence rate $\sigtd$, 
\begin{align}
  \sigtd &=  \frac{2ug}{\C} + \frac{ \phi}{{2}ug} + O\left(\frac{\phi^2 }{(ug)^3} \right).
\end{align}
However, the adaptive term remains subleading to the housekeeping term for large
$g$; this is true even if we assume that $\phi$ is proportional to $g$.
Hence, the total interference load, $\L_{\rm int} = g \sigtd =2 ug^2 / c + \phi
/ u + \dots$, retains the leading nonlinearity generated by housekeeping
evolution, as given by Eq.~(\ref{load}).
Only for very high fitness flux ($\phi \gg u^2g^2/c$), coalescence becomes dominated by adaptation, leading to a substantial decrease in the efficacy of selection.

\section{Numerical simulations of phenotypic interference}
\paragraph{In-silico evolution of stability traits.} 
We use a  Wright-Fisher process to simulate the evolution of stability traits in a population.
A population consists of $N$ individuals  with genomes $\a^{(1)}, \dots,
\a^{(N)}$. A genotype  $\a =
(\a_1, \dots, \a_g)$ consists of $g$ segments;  each segment is a subsequence $\a_i = (a_{i,1}, \dots, a_{i,\ell})$ with binary alleles $a_{j,k} = 0,1$  ($i = 1, \dots, g$; $k = 1, \dots, \ell$). A segment $\a$ defines a stability trait $G (\a) = 
\sum_{k =1}^\ell \cE_k a_k + G_0$, where $G_0$ is the expectation value of the trait under neutral evolution.
The resulting effect distribution of point mutations has as a second moment $\epsilon_{G}^2 =  \sum_{\i =
  1}^\ell \cE_k^2 / \ell$ and a first moment 
$\kappa_0 \epsilon_{G} = \sum_{\i =
  1}^\ell \cE_k (1 - 2\langle a_k \rangle )/\ell$,
where $\langle a_{k} \rangle$ is the state-dependent probability of a mutation at site $k$ being beneficial and brackets $\langle. \rangle$ denote averaging across parallel simulations or time.
The genomic fitness is  $f(\a)=\sum_{i=1}^g f(G(\a_i); f_{0,i})$ with $f(G)$ given by Eq.~(\ref{eq:fG}) and gene-specific amplitudes $f_{0,i}$. 
In each generation, the sequences undergo point mutations with probability $
\mu \tau_0$ for each site, where $\tau_0$ is the generation time, and 
the sequences of the next generation are drawn by multinomial sampling
with a probabilities proportional to $1 + \tau_0 f(\a)$. 

Simulations are performed with parameters $N=1000$, $N\mu=0.0125$,  each trait with genomic base of size $\ell = 100$, and each site with equal effect $\cE_{k} = 1$. The quantitative trait dynamics is insensitive to the form of the effect distribution~\cite{Nourmohammad:2013ty,Held2014}. To increase the performance of the simulations, we do not keep track of the full genome. We only store the number of deleterious alleles  $n_{i}=\sum_{k=1}^{\ell} a_{i,k}$ for each trait, we draw mutations with rate $u=\mu\ell$, and we assign to each mutation a beneficial change $\cE$ with probability $n_{i}/\ell$ and a deleterious change~$-\cE$ otherwise. This procedure produces the correct genome statistics for bi-allelic sites with uniform trait effects $\cE_i = \cE$. Simulation data are shown with theory curves for $\kappa = 1$, which provide a good fit to all amplitudes; the input $\kappa_0$ is different by a factor of order 1 which includes fluctuation effects (Section~2).

\paragraph{Housekeeping evolution.}
For the simulations in Figs.~2 and~3A, where we are not explicitly interested in the loss of
genes, we use an exponential approximation of the stable regime of the stability fitness landscape. The reason is a limited  accessible parameter range in 
simulations constraining the values of $f_0$ and $\sigtd$ due
to finite $N$. We checked that the exponential approximation 
gives the same results as the full model in the regime $f_0 / \sigtd \gg 1$, where the gene loss rate in the biophysical
landscape is negligible.

\paragraph{Loss rate measurements.}
In the biophysical landscape used in Fig~3B, a long-term stationary population is maintained by evolving  70\% of the traits in a biophysical fitness landscape with selection $f_0$; the remaining 30\% of the traits are modeled to be essential  with selection $10f_0$. Gene loss is defined by the condition  $\DG < -3.5 \kbT$. To maintain a constant number of genes, lost genes are replaced immediately with an input trait value $G > 0$. 

\paragraph{Recombination.}
For simulations with recombination (Fig.~3C), we draw recombination events with rate $N R$ for the whole population from a Poisson distribution.
Each recombination event is implemented as one crossover between the genomes of
two individuals at a random, uniformly distributed position of the genomes.

\clearpage

\begin{figure}
  \includegraphics[width=\textwidth]{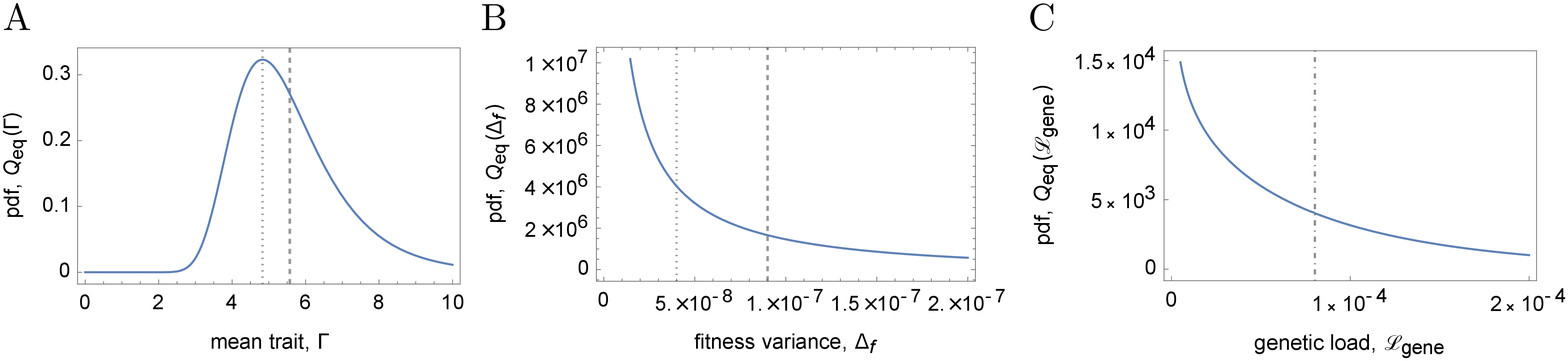}
  \caption{{\bf Equilibrium distributions under stochastic evolution.} The figure shows the probability density functions (A) of the mean population trait, $Q_{\rm eq} (\Gamma)$, (B) of the conditional expected fitness variance, $Q_{\rm eq}(\flux)$, and (C) of the genetic load per gene, $Q_{\rm eq} (\L_{\rm gene})$; see Eqs.~(\ref{QGamma}) -- (\ref{Qf}). These distributions measure deviations from long-term averages (dashed lines), which are generated by genetic drift and draft. The corresponding deterministic solutions are marked by dotted lines; both lines coincide in (C). All pdfs are shown for $\sigtd=f_{0}/100=10^{-4}$; other parameters as in Fig.~2.  See section~2 of S1 Appendix. }
\label{S1_Fig}
\end{figure}

\begin{figure}
\centering
\includegraphics[width=0.5\textwidth]{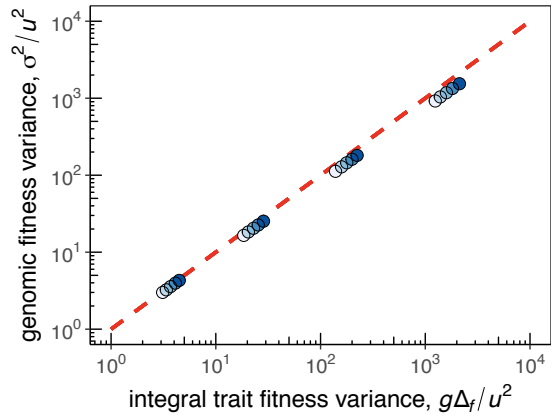}

  \caption{{\bf Additivity of the genomic fitness variance.} 
For housekeeping evolution in the minimal biophysical model, we plot the total fitness variance, $\sigma^2$, against the additive part $\Delta_{f,1} + \dots + \Delta_{f,g}$. The additivity is used in the closure of the evolutionary dynamics, Eqs.~(2)~--~(4). 
  }
\label{S2_Fig}
\end{figure}

\begin{figure}
  \includegraphics[width=\textwidth]{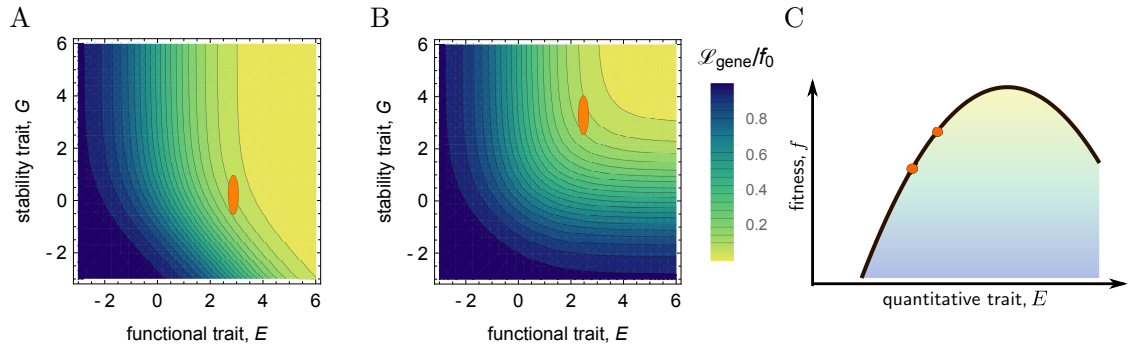}

  \caption{{\bf Extended fitness landscapes.} 
 (A, B) Thermodynamic fitness landscapes $f(G,E)$ of the stability-affinity model, Eqs.~(\ref{p++1}) -- (\ref{fEG}), are shown as functions of the stability $G$ and the affinity $E$.  Stable populations, characterized by stationary mean values $(\Gamma_G, \Gamma_E)$ and variances $(\Delta_G, \Delta_E)$, are marked by red ellipsoids. (A) Thermodynamic equilibrium. (B) Non-equilibrium driven by active degradation of folded proteins. In the high-fitness part, this landscape becomes approximately additive in $G$ and $E$. (C) Quadratic fitness landscape $f(E)$, Eq.~(\ref{eq:fquad}), as a minimal model for stabilizing selection on a quantitative trait $E$. Stable population states on a flank of the landscape are marked by red dots. See section~3 of S1 Appendix. 
  }
\label{S3_Fig}
\end{figure}

\clearpage

\begin{table}[th!]
\centering
\begin{tabular}{c||l|l|l}
 \multicolumn{1}{l||}{}  &%
   \multicolumn{ 1}{c|}{{\em Saccharomyces cerevisiae}} & \multicolumn{ 1}{c|}{{\em Drosophila melanogaster}} & \multicolumn{ 1}{c}{{\em Arabidopsis thaliana}} \\ \hline \hline
\multicolumn{1}{c||}{$\mu$} & %
$3\!\cdot\!10^{-8}$  \cite{Lynch08072008} & $3\!\cdot\!10^{-9}$  \cite{Keightley2014}  & $7\!\cdot\!10^{-9} $\cite{Ossowski92} \\ \hline
\multicolumn{1}{c||}{$\ell$} &%
 $1401$  \cite{Harrison01022003} & $1500 $ \cite{Harrison01022003} & $2232$  \cite{Derelle2006} \\ \hline 
\multicolumn{1}{c||}{$g$}   & %
$6563$  \cite{Derelle2006} & $14332$ \cite{Harrison01022003} & $26990$  \cite{B903031J} \\ \hline 
\multicolumn{1}{c||}{$R$}  & %
 $3\!\cdot\!10^{-2}$ \cite{Ruderfer2006} & $2\!\cdot\!10^{-3}\-- 1\!\cdot\! 10^{-0}$ \cite{FistonLavier201018,Comeron2012} & $2\!\cdot\!10^{-0}$  \cite{Salome2012} \\ \hline
\multicolumn{1}{c||}{${R}^{*}$}  & %
 $6\!\cdot\!10^{-5}$  & $1\!\cdot\!10^{-3}$ & $9\!\cdot\!10^{-3}$
\end{tabular}
\caption{ {\bf Genome data and estimates of threshold recombination rates.}
  Point mutation rate $\mu$, average gene length $\ell$ (in bp), gene number $g$
  and recombination rate $R$ per genome (map length) are shown for three
  recombining species. The parameter range for \emph{D. melanogaster} describes
  local recombination rates in different parts of the chromosomes (in the same
  units) \cite{Comeron2012}. An upper bond of the threshold recombination rate
  $R^*$ marking the transition to sexual evolution is obtained from Eq.~(8)
  (with $ug=\mu\ell g$ and $c \approx c_0\approx 100$).}
\label{S1_Table}
\end{table}


\begin{thebibliography}{10}
\setcounter{firstbib}{\value{enumiv}}
\bibitem{Wiser2013}
Wiser MJ, Ribeck N, Lenski RE.
\newblock Long-Term Dynamics of Adaptation in Asexual Populations.
\newblock Science. 2013;342(6164):1364--1367.
\newblock doi:{10.1126/science.1243357}.

\bibitem{Barroso-Batista2014}
Barroso-Batista J, Sousa A, Louren{\c c}o M, Bergman ML, Sobral D, Demengeot J,
  et~al.
\newblock The First Steps of Adaptation of \emph{{E}scherichia coli} to the Gut
  Are Dominated by Soft Sweeps.
\newblock PLoS Genet. 2014;10(3):e1004182.
\newblock doi:{10.1371/journal.pgen.1004182}.

\bibitem{Betancourt2009}
Betancourt AJ, Welch JJ, Charlesworth B.
\newblock Reduced Effectiveness of Selection Caused by a Lack of Recombination.
\newblock Current Biology. 2009;19(8):655--660.
\newblock doi:{10.1016/j.cub.2009.02.039}.

\bibitem{Strelkowa2012}
Strelkowa N, L{\"a}ssig M.
\newblock Clonal Interference in the Evolution of Influenza.
\newblock Genetics. 2012;192(2):671--682.
\newblock doi:{10.1534/genetics.112.143396}.

\bibitem{Gerrish1998}
Gerrish PJ, Lenski RE.
\newblock The fate of competing beneficial mutations in an asexual population.
\newblock Genetica. 1998;102:127--144.
\newblock doi:{10.1023/A:1017067816551}.

\bibitem{Desai01072007}
Desai MM, Fisher DS.
\newblock Beneficial Mutation--Selection Balance and the Effect of Linkage on
  Positive Selection.
\newblock Genetics. 2007;176(3):1759--1798.
\newblock doi:{10.1534/genetics.106.067678}.

\bibitem{Rouzine200824}
Rouzine IM, Brunet {\'E}, Wilke CO.
\newblock The traveling-wave approach to asexual evolution: Muller's ratchet
  and speed of adaptation.
\newblock Theor Popul Biol. 2008;73(1):24 -- 46.
\newblock doi:{10.1016/j.tpb.2007.10.004}.

\bibitem{Hallatschek01022011}
Hallatschek O.
\newblock The noisy edge of traveling waves.
\newblock Proc Natl Acad Sci. 2011;108(5):1783--1787.
\newblock doi:{10.1073/pnas.1013529108}.

\bibitem{schiffels2011emergent}
Schiffels S, Sz{\"o}ll{\H{o}}si GJ, Mustonen V, L{\"a}ssig M.
\newblock Emergent neutrality in adaptive asexual evolution.
\newblock Genetics. 2011;189(4):1361--1375.
\newblock doi:{10.1534/genetics.111.132027}.

\bibitem{Good2012}
Good BH, Rouzine IM, Balick DJ, Hallatschek O, Desai MM.
\newblock Distribution of fixed beneficial mutations and the rate of adaptation
  in asexual populations.
\newblock Proc Natl Acad Sci. 2012;109(13):4950--4955.
\newblock doi:{10.1073/pnas.1119910109}.

\bibitem{Neher08012013}
Neher RA, Hallatschek O.
\newblock Genealogies of rapidly adapting populations.
\newblock Proc Natl Acad Sci. 2013;110(2):437--442.
\newblock doi:{10.1073/pnas.1213113110}.

\bibitem{Neher2013}
Neher RA, Kessinger TA, Shraiman BI.
\newblock Coalescence and genetic diversity in sexual populations under
  selection.
\newblock Proc Natl Acad Sci. 2013;110(39):15836--15841.
\newblock doi:{10.1073/pnas.1309697110}.

\bibitem{Visser1999}
de~Visser AGJM, Zeyl CW, Gerrish PJ, Blanchard JL, Lenski RE.
\newblock Diminishing Returns from Mutation Supply Rate in Asexual Populations.
\newblock Science. 1999;283(5400):404--406.
\newblock doi:{10.1126/science.283.5400.404}.

\bibitem{Perfeito2007}
Perfeito L, Fernandes L, Mota C, Gordo I.
\newblock Adaptive Mutations in Bacteria: High Rate and Small Effects.
\newblock Science. 2007;317(5839):813--815.
\newblock doi:{10.1126/science.1142284}.

\bibitem{McDonald2016}
McDonald MJ, Rice DP, Desai MM.
\newblock Sex speeds adaptation by altering the dynamics of molecular
  evolution.
\newblock Nature. 2016;531:233--236.
\newblock doi:{10.1038/nature17143}.

\bibitem{Cooper2007}
Cooper TF.
\newblock Recombination Speeds Adaptation by Reducing Competition between
  Beneficial Mutations in Populations of { \em {E}scherichia coli}.
\newblock PLoS Biol. 2007;5(9):e225.
\newblock doi:{10.1371/journal.pbio.0050225}.

\bibitem{Couce2017}
Couce A, Caudwell LV, Feinauer C, Hindr{\'e} T, Feugeas JP, Weigt M, et~al.
\newblock Mutator genomes decay, despite sustained fitness gains, in a
  long-term experiment with bacteria.
\newblock Proc Natl Acad Sci. 2017;114(43):E9026--E9035.
\newblock doi:{10.1073/pnas.1705887114}.

\bibitem{Fisher1930}
Fisher RA.
\newblock The genetical theory of natural selection.
\newblock Oxford Clarendon Press; 1930.

\bibitem{Muller1932}
Muller HJ.
\newblock Some Genetic Aspects of Sex.
\newblock Am Nat. 1932;66(703):118--138.
\newblock doi:{10.1086/280418}.

\bibitem{Eigen1971}
Eigen M.
\newblock Selforganization of matter and the evolution of biological
  macromolecules.
\newblock Naturwissenschaften. 1971;58(10):465--523.
\newblock doi:{10.1007/BF00623322}.

\bibitem{Felsenstein1974}
Felsenstein J.
\newblock The Evolutionary Advantage of Recombination.
\newblock Genetics. 1974;78(2):737--756.

\bibitem{Kondrashov1993}
Kondrashov AS.
\newblock Classification of hypotheses on the advantage of amphimixis.
\newblock J Hered. 1993;84(5):372--387.
\newblock doi:{10.1093/oxfordjournals.jhered.a111358}.

\bibitem{Gerland2002}
Gerland U, Hwa T.
\newblock {On the selection and evolution of regulatory DNA motifs}.
\newblock J Mol Evol. 2002;55(4):386--400.
\newblock doi:{10.1007/s00239-002-2335-z}.

\bibitem{Berg2004}
Berg J, Willmann S, L\"assig M.
\newblock Adaptive evolution of transcription factor binding sites.
\newblock BMC Evol Biol. 2004;4(1):42.
\newblock doi:{10.1186/1471-2148-4-42}.

\bibitem{Chen01102009}
Chen P, Shakhnovich EI.
\newblock Lethal Mutagenesis in Viruses and Bacteria.
\newblock Genetics. 2009;183(2):639--650.
\newblock doi:{10.1534/genetics.109.106492}.

\bibitem{Goldstein2011}
Goldstein RA.
\newblock The evolution and evolutionary consequences of marginal
  thermostability in proteins.
\newblock Proteins: Struct , Funct , Bioinf. 2011;79(5):1396--1407.
\newblock doi:{10.1002/prot.22964}.

\bibitem{Serohijos201484}
Serohijos AWR, Shakhnovich EI.
\newblock Merging molecular mechanism and evolution: theory and computation at
  the interface of biophysics and evolutionary population genetics.
\newblock Curr Opin Struct Biol. 2014;26(0):84--91.
\newblock doi:{10.1016/j.sbi.2014.05.005}.

\bibitem{Manhart2015}
Manhart M, Morozov AV.
\newblock Protein folding and binding can emerge as evolutionary spandrels
  through structural coupling.
\newblock Proc Natl Acad Sci. 2015;112(6):1797--1802.
\newblock doi:{10.1073/pnas.1415895112}.

\bibitem{Chi2016}
Chi PB, Liberles DA.
\newblock Selection on protein structure, interaction, and sequence.
\newblock Protein Sci. 2016;25(7):1168--1178.
\newblock doi:{10.1002/pro.2886}.

\bibitem{Muller1964}
Muller HJ.
\newblock The relation of recombination to mutational advance.
\newblock Mutat Res. 1964;106:2--9.
\newblock doi:{10.1016/0027-5107(64)90047-8}.

\bibitem{Gordo2000}
Gordo I, Charlesworth B.
\newblock The degeneration of asexual haploid populations and the speed of
  {M}uller's ratchet.
\newblock Genetics. 2000;154(3):1379--1387.

\bibitem{Gillespie909}
Gillespie JH.
\newblock Genetic Drift in an Infinite Population: The Pseudohitchhiking Model.
\newblock Genetics. 2000;155(2):909--919.

\bibitem{Good2014}
Good BH, Walczak AM, Neher RA, Desai MM.
\newblock Genetic Diversity in the Interference Selection Limit.
\newblock PLoS Genet. 2014;10(3):e1004222.
\newblock doi:{10.1371/journal.pgen.1004222}.

\bibitem{Rice01052015}
Rice DP, Good BH, Desai MM.
\newblock The Evolutionarily Stable Distribution of Fitness Effects.
\newblock Genetics. 2015;200(1):321--329.
\newblock doi:{10.1534/genetics.114.173815}.

\bibitem{KlemmerLaessig2018}
Klemmer D, L{\"a}ssig M; 2018.

\bibitem{Hochstrasser1996}
Hochstrasser M.
\newblock Ubiquitin-dependent protein degradation.
\newblock Annu Rev Genet. 1996;30:405--439.
\newblock doi:{10.1146/annurev.genet.30.1.405}.

\bibitem{Cheron2016}
Ch{\'e}ron N, Serohijos AWR, Choi JM, Shakhnovich EI.
\newblock Evolutionary dynamics of viral escape under antibodies stress: A
  biophysical model.
\newblock Protein Sci. 2016;25(7):1332--1340.
\newblock doi:{10.1002/pro.2915}.

\bibitem{Scott2010}
Scott M, Gunderson CW, Mateescu EM, Zhang Z, Hwa T.
\newblock Interdependence of Cell Growth and Gene Expression: Origins and
  Consequences.
\newblock Science. 2010;330(6007):1099--1102.
\newblock doi:{10.1126/science.1192588}.

\bibitem{Lynch2015}
Lynch M, Marinov GK.
\newblock The bioenergetic costs of a gene.
\newblock Proc Natl Acad Sci. 2015;112(51):15690--15695.
\newblock doi:{10.1073/pnas.1514974112}.

\bibitem{Weissman2012}
Weissman DB, Barton NH.
\newblock Limits to the Rate of Adaptive Substitution in Sexual Populations.
\newblock PLoS Genet. 2012;8(6):1--18.
\newblock doi:{10.1371/journal.pgen.1002740}.

\bibitem{Neher2013a}
Neher RA.
\newblock Genetic Draft, Selective Interference, and Population Genetics of
  Rapid Adaptation.
\newblock Annu Rev Ecol Evol Syst. 2013;44(1):195--215.
\newblock doi:{10.1146/annurev-ecolsys-110512-135920}.

\bibitem{Comeron2012}
Comeron JM, Ratnappan R, Bailin S.
\newblock The Many Landscapes of Recombination in{ \em Drosophila
  melanogaster}.
\newblock PLoS Genet. 2012;8(10):1--21.
\newblock doi:{10.1371/journal.pgen.1002905}.

\bibitem{Schiffels2017}
{Schiffels} S, {Mustonen} V, {L{\"a}ssig} M.
\newblock {The asexual genome of Drosophila}.
\newblock ArXiv e-prints. 2017;(1711.10849).

\bibitem{Mustonen2008}
Mustonen V, Kinney J, Callan CGJ, L{\"a}ssig M.
\newblock {Energy-dependent fitness: A quantitative model for the evolution of
  yeast transcription factor binding sites}.
\newblock Proc Natl Acad Sci USA. 2008;105(34):12376--12381.
\newblock doi:{10.1073/pnas.0805909105}.

\bibitem{Lassig2007}
L{\"a}ssig M.
\newblock {From biophysics to evolutionary genetics: statistical aspects of
  gene regulation}.
\newblock BMC Bioinformatics. 2007;8(Suppl 6):S7.
\newblock doi:{10.1186/1471-2105-8-s6-s7}.

\bibitem{Nourmohammad:2013ty}
Nourmohammad A, Schiffels S, L{\"a}ssig M.
\newblock {Evolution of molecular phenotypes under stabilizing selection}.
\newblock J Stat Mech Theor Exp. 2013;2013(01):P01012.
\newblock doi:{10.1088/1742-5468/2013/01/P01012}.

\bibitem{Tokuriki2007}
Tokuriki N, Stricher F, Schymkowitz J, Serrano L, Tawfik DS.
\newblock The Stability Effects of Protein Mutations Appear to be Universally
  Distributed.
\newblock J Mol Biol. 2007;369(5):1318--1332.
\newblock doi:{10.1016/j.jmb.2007.03.069}.

\bibitem{Zeldovich07}
Zeldovich KB, Chen P, Shakhnovich EI.
\newblock Protein stability imposes limits on organism complexity and speed of
  molecular evolution.
\newblock Proc Natl Acad Sci. 2007;104(41):16152--16157.
\newblock doi:{10.1073/pnas.0705366104}.

\bibitem{Kinney2010}
Kinney JB, Murugan A, Callan CG, Cox EC.
\newblock Using deep sequencing to characterize the biophysical mechanism of a
  transcriptional regulatory sequence.
\newblock Proc Natl Acad Sci. 2010;107(20):9158--9163.
\newblock doi:{10.1073/pnas.1004290107}.

\bibitem{Tugrul2015}
Tu\u{g}rul M, Paix\~{a}o T, Barton NH, Tka\v{c}ik G.
\newblock Dynamics of Transcription Factor Binding Site Evolution.
\newblock PLoS Genet. 2015;11(11):1--28.
\newblock doi:{10.1371/journal.pgen.1005639}.
\setcounter{firstbib}{\value{enumiv}}
\end{thebibliography}

\begin{thebibliography}{10}
\setcounter{enumiv}{\value{firstbib}}
\bibitem{Phillips2013}
Phillips R, Kondev J, Theriot J, Orme N.
\newblock Physical Biology of the Cell.
\newblock Garland Science; 2013.

\bibitem{Monod1965}
Monod J, Wyman J, Changeux JP.
\newblock On the nature of allosteric transitions: A plausible model.
\newblock Journal of Molecular Biology. 1965;12(1):88--118.
\newblock doi:{10.1016/s0022-2836(65)80285-6}.

\bibitem{Otwinowski2018}
Otwinowski J.
\newblock Inferring protein stability and function from a high-throughput
  assay.
\newblock arXiv preprint. 2018;1802.08744.

\bibitem{Lande1976}
Lande R.
\newblock Natural selection and random genetic drift in phenotypic evolution.
\newblock Evolution. 1976;30(2):314--334.

\bibitem{deVladar2014}
{{de Vladar}} HP, Barton N.
\newblock Stability and Response of Polygenic Traits to Stabilizing Selection
  and Mutation.
\newblock Genetics. 2014;197(2):749--767.
\newblock doi:{10.1534/genetics.113.159111}.

\bibitem{nourmohammad2013universality}
Nourmohammad A, Held T, L{\"a}ssig M.
\newblock Universality and predictability in molecular quantitative genetics.
\newblock Curr Opin Genet Dev. 2013;23(6):684--693.
\newblock doi:{10.1016/j.gde.2013.11.001}.

\bibitem{Mustonen2010}
Mustonen V, L{\"a}ssig M.
\newblock Fitness flux and ubiquity of adaptive evolution.
\newblock Proc Natl Acad Sci. 2010;107:4248--53.
\newblock doi:{10.1073/pnas.0907953107}.

\bibitem{Held2014}
Held T, Nourmohammad A, L{\"a}ssig M.
\newblock Adaptive evolution of molecular phenotypes.
\newblock J Stat Mech: Theory Exp. 2014;2014(9):P09029.
\newblock doi:{10.1088/1742-5468/2014/09/p09029}.

\bibitem{Lynch08072008}
Lynch M, Sung W, Morris K, Coffey N, Landry CR, Dopman EB, et~al.
\newblock A genome-wide view of the spectrum of spontaneous mutations in yeast.
\newblock Proc Natl Acad Sci. 2008;105(27):9272--9277.
\newblock doi:{10.1073/pnas.0803466105}.

\bibitem{Keightley2014}
Keightley PD, Ness RW, Halligan DL, Haddrill PR.
\newblock Estimation of the Spontaneous Mutation Rate per Nucleotide Site in a
  Drosophila melanogaster Full-Sib Family.
\newblock Genetics. 2014;196(1):313--320.
\newblock doi:{10.1534/genetics.113.158758}.

\bibitem{Ossowski92}
Ossowski S, Schneeberger K, Lucas-Lled{\'o} JI, Warthmann N, Clark RM, Shaw RG,
  et~al.
\newblock The Rate and Molecular Spectrum of Spontaneous Mutations in{ \em
  Arabidopsis thaliana}.
\newblock Science. 2010;327(5961):92--94.
\newblock doi:{10.1126/science.1180677}.

\bibitem{Harrison01022003}
Harrison PM, Milburn D, Zhang Z, Bertone P, Gerstein M.
\newblock Identification of pseudogenes in the { \em Drosophila melanogaster}
  genome.
\newblock Nucleic Acids Res. 2003;31(3):1033--1037.
\newblock doi:{10.1093/nar/gkg169}.

\bibitem{Derelle2006}
Derelle E, Ferraz C, Rombauts S, Rouz\'{e} P, Worden AZ, Robbens S, et~al.
\newblock Genome analysis of the smallest free-living eukaryote { \em
  Ostreococcus tauri} unveils many unique features.
\newblock Proc Natl Acad Sci. 2006;103(31):11647--11652.
\newblock doi:{10.1073/pnas.0604795103}.

\bibitem{B903031J}
Feng J, Naiman DQ, Cooper B.
\newblock Coding {DNA} repeated throughout intergenic regions of the { \em
  {A}rabidopsis thaliana} genome: evolutionary footprints of {RNA} silencing.
\newblock Mol Biosyst. 2009;5:1679--1687.
\newblock doi:{10.1039/B903031J}.

\bibitem{Ruderfer2006}
Ruderfer DM, Pratt SC, Seidel HS, Kruglyak L.
\newblock Population genomic analysis of outcrossing and recombination in
  yeast.
\newblock Nat Genet. 2006;38(9):1077--1081.
\newblock doi:{10.1038/ng1859}.

\bibitem{FistonLavier201018}
Fiston-Lavier AS, Singh ND, Lipatov M, Petrov DA.
\newblock {\em Drosophila melanogaster} recombination rate calculator.
\newblock Gene. 2010;463(1–2):18 -- 20.
\newblock doi:{10.1016/j.gene.2010.04.015}.

\bibitem{Salome2012}
Salom\'{e} PA, Bomblies K, Fitz J, Laitinen RAE, Warthmann N, Yant L, et~al.
\newblock The recombination landscape in { \em Arabidopsis thaliana} {F$_2$}
  populations.
\newblock Heredity. 2012;108(4):447--455.
\newblock doi:{10.1038/hdy.2011.95}.
\end{thebibliography}
\end{document}